\documentclass[10pt,twocolumn,letterpaper]{article}

\usepackage{cvpr}              

\usepackage{graphicx}
\usepackage{amsmath}
\usepackage{amssymb}
\usepackage{booktabs}
\usepackage{algorithm,algorithmic}

\usepackage[pagebackref,breaklinks,colorlinks]{hyperref}

\usepackage[capitalize]{cleveref}
\crefname{section}{Sec.}{Secs.}
\Crefname{section}{Section}{Sections}
\Crefname{table}{Table}{Tables}
\crefname{table}{Tab.}{Tabs.}

\def\cvprPaperID{*****} 
\def\confName{CVPR}
\def\confYear{2022}

\begin{document}

\title{Hadamard Row-Wise Generation Algorithm}

\author{Brayan Monroy \& Jorge Bacca\\
Universidad Industrial de Santander, Bucaramanga, Colombia\\
\url{https://github.com/bemc22/hadamard-spc}
}
\maketitle

\begin{abstract}
We present an efficient algorithm for generating specific Hadamard rows, addressing the memory demands of pre-computing the entire matrix. Leveraging Sylvester's recursive construction, our method generates the required $i$-th row on demand, significantly reducing computational resources. The algorithm uses the Kronecker product to construct the desired row from the binary representation of the index without creating the full matrix. This approach is particularly useful for single-pixel imaging systems that need only one row at a time.  
\end{abstract} \vspace{-1em}

\section{Method}
\label{sec:intro}

Computing the $i$-th row $\textbf{h}_i$ of a Hadamard matrix $\textbf{H}$ of order $2^n$ usually requires pre-computing the entire matrix. This process can be memory-intensive, particularly when $n$ is large. However, in certain applications, such as Hadamard single-pixel imaging, only individual Hadamard rows are needed at any given time~\cite{duarte2008single}. To address this, we have developed an algorithm for row-wise generation that calculates the specific coefficients of the $i$-th row without generating the entire matrix. Specifically, following Sylvester's construction, a Hadamard matrix of order $2^n$ can be recursively constructed from a base matrix of order two and Kronecker products as follows \begin{equation}
    \textbf{H}_{2^n} = 
    \begin{bmatrix}
     \textbf{H}_{2^{n-1}} & \textbf{H}_{2^{n-1}} \\ \textbf{H}_{2^{n-1}} & - \textbf{H}_{2^{n-1}}
    \end{bmatrix} = \textbf{H}_2 \otimes \textbf{H}_{2^{n-1}},
\end{equation} with $2 \leq n \in \mathbb{N}$ where $\otimes$ denotes the Kronecker product. Hence, each row of the Hadamard matrix of order $2^n$ can be expressed as the Kronecker product of the first or second row of the Hadamard matrix of order 2. The sequence of Kronecker products is derived from the binary representation $i_{10} = (b_nb_{n-1}\dots b_1 b_0)_2$ of the $i$-th row as follows

\begin{equation}
    \textbf{h}_i = \bigotimes_{k=n}^0 \textbf{h}_{b_k} = \textbf{h}_{b_n} \otimes \textbf{h}_{b_{n-1}} \otimes \cdots  \otimes \textbf{h}_{b_1} \otimes \textbf{h}_{b_0}
\end{equation}

\begin{figure*}[!t]
    \centering
    \includegraphics[width=\linewidth]{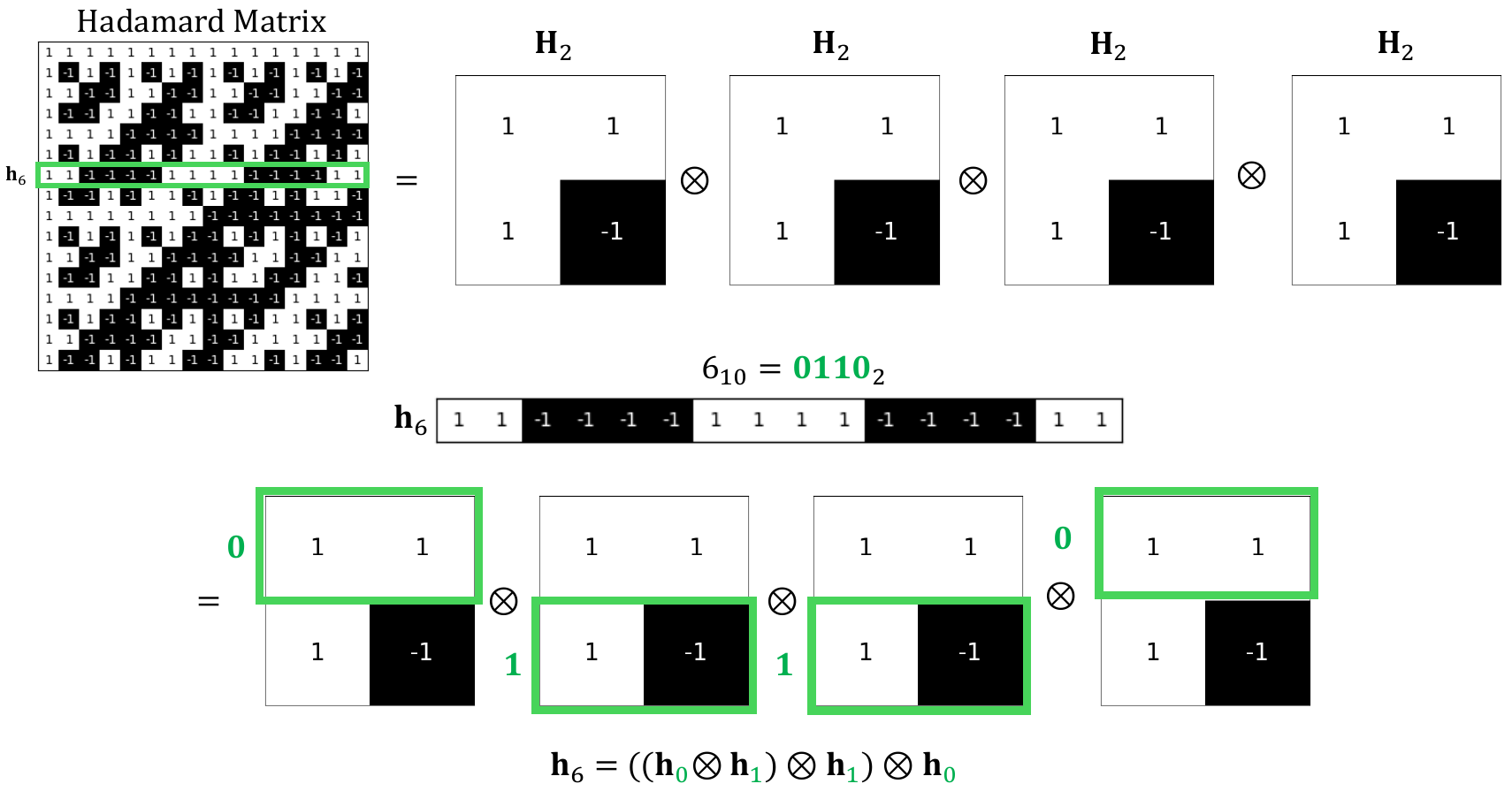}
    \caption{Hadamard Row-Wise Generation Algorithm. In the case of $\textbf{h}_6$, the 6-th index has a binary representation of $6_{10} = \textbf{0110}_2 $. The digits in this binary representation can be used to index the Kronecker product of $n$, 2-order Hadamard matrices, where 0/1 corresponds to using the first or second row, respectively.}
    \label{fig:visual}
\end{figure*}

In this context, we present the Algorithm~\ref{alg:row_wise}, which involves setting the base 2-order Hadamard matrix and mapping the specified index to its binary representation. An iterative loop processes the binary digits to select the appropriate rows of the 2-order Hadamard matrix, ultimately constructing the desired Hadamard row via the cumulative Kronecker product as presented in Figure~\ref{fig:visual}. Additionally, this algorithm can be adapted for other Hadamard ordering strategies, as it relies mainly on permuting ordering indexes. The code implementation is available on GitHub.

It is important to note that the generation of a Hadamard matrix row without constructing the entire matrix is less discussed in the literature, although it is a natural extension of their recursive nature, as proposed in~\cite{sylvester1867lx}. We believe that the algorithm provided contributes to a more detailed documentation of this strategy.

\textit{Computational complexity. } The computational complexity of Algorithm~\ref{alg:row_wise} relies on the complexity in compute $n$ times Kronecker product of vector of size 2. In this sense, for each Kronecker product between the vector of size $2^{k-1}$ (resulting from the previous $k-1$ Kronecker products) and a vector of size 2, the number of multiplication required is $2 \times 2^{k-1} = 2^k$. Thus, the total computational complexity $\mathcal{C}(n)$ for performing $n$ Kronecker products is the sum of the number of multiplications for each step
\begin{equation}
    \mathcal{C}(n) = 2^1 + 2^2 + 2^3 +  \cdots + 2^n = \sum_{k=1}^n 2^k,
\end{equation}
which consists of a geometric series that can be simplified as $\mathcal{C}(n) = 2^{n+1} - 2$, with the dominant term in $\mathcal{C}(n)$ being $2^{n+1}$, so the asymptotic computational complexity of Algorithm~\ref{alg:row_wise} is $\mathcal{O}(2^{n+1}) \sim \mathcal{O}(2^{n})$.

\begin{algorithm}[H]  
 \caption{Hadamard Row-Wise Generation} \label{alg:row_wise}
 \begin{algorithmic}[1]
 \renewcommand{\algorithmicrequire}{\textbf{Input:}}
 \renewcommand{\algorithmicensure}{\textbf{Output:}}
 \REQUIRE $i$, $n$, $\textbf{H}_2$
  \STATE Set $indxs = \texttt{dec2bin}(i, n) $
  \STATE Set $\textbf{h}_i = 1$
  \FOR {$j = 0$ to $\texttt{length}(indxs)$}
  \STATE Set $\textbf{h}_{temp} = \textbf{H}_2[ indxs[j], :]$
  \STATE Set $\textbf{h}_i = \texttt{kron}(\textbf{h}_i, \textbf{h}_{temp})$
  \ENDFOR
 \RETURN  $\textbf{h}_i$
 \end{algorithmic} 
 \end{algorithm}

{\small
\bibliographystyle{ieee_fullname}
\bibliography{egbib}
}

\end{document}